\documentclass{pasj00}
\draft

\begin{document}
\SetRunningHead{S. Kato}
               {Excitation of Disk Oscillations in Deformed Disks}
\Received{2011/0/00}
\Accepted{2011/0/00}

\title{Resonant Excitation of Disk Oscillations in Deformed Disks V:  
        Effects of Dissipative Process}

\author{Shoji \textsc{Kato}}%
\affil{2-2-2 Shikanodai-Nishi, Ikoma-shi, Nara 630-0114}
\email{kato.shoji@gmail.com; kato@kusastro.kyoto-u.ac.jp}


%


\maketitle

\begin{abstract}
It is suggested that a set of positive- and negative-energy oscillations can be resonantly 
excited in the inner region of deformed (warped or eccentric) relativistic disks.
In this paper we examine how a dissipative process affects on this wave excitation process.
The results show that when the resonant condition in frequency is roughly satisfied and thus the 
oscillations are excited, introduction of a 
dissipative process works so as to decrease the growth rate of the oscillations.
When the frequency difference of the two oscillations deviates more than a certain amount from 
that required by resonant condition, however,
the oscillations are excited by introduction of dissipative process. 
This excitation by dissipative process can be understood as a special example of the double-diffusive instability. 
\end{abstract}


\section{Introduction}

Investigation of the origin of high-frequency quasi-periodic oscillations (HF QPOs)
observed in neutron-star and black-hole low-mass X-ray binaries (LMXBs) is one of important subjects of astrophysical
disks, since it will give clues to understand the structure of the innermost region of relativistic disks, 
and to estimate spins of the central sources.
One of possible causes of HF QPOs is disk oscillations in relativistic disks [see Kato (2001) or Kato et al. (2008)
for review of disk oscillations in relativistic disks].
Related to this subject, Kato (2004, 2008a, b) suggested that a pair of positive- and negative-
energy oscillations can be excited in deformed (warped or eccentric) disks by their resonant interaction
through disk deformation.
The presence of this excitation mechanism seems to be confirmed by numerical calculations
(Ferreira and Ogilvie 2008; Oktariani et al. 2010) and by analytical examinations (Kato et al. 2011) made by a different
method from Kato's original one.

There remains, however, an important disagreement between analytical results by Kato' group (Kato 2004, 2008a, b; Kato et al. 2011)
and numerical ones by Ferreira and Ogilvie (2008).
In Ferreira and Ogilvie's ones, a dissipative process acting on one of oscillations  
is necessary to excite the pair of the oscillations, while in Kato's ones
any energy dissipative process is unnecessary to excite the set of positive- and negative-energy oscillations.

Let us consider two oscillation modes the set of whose frequencies and azimuthal wave numbers are, respectively, 
($\omega_1$, $m_1$) and ($\omega_2$, $m_2$).
Disk deformation is denoted by ($\omega_{\rm D}$, $m_{\rm D}$), and $\omega_{\rm D}$ is fixed in the 
present problem.
In the case where the disk deformation is a stationary warp, for example, $\omega_{\rm D}=0$ and
$m_{\rm D}=1$.
Necessary conditions of resonant interaction between the above two oscillations through
the disk deformation are 
\begin{equation}
       m_2=m_1\pm m_{\rm D} \quad {\rm and} \quad \omega_2\sim \omega_1\pm \omega_{\rm D}.
\label{0}
\end{equation}
In Kato's original studies (2004, 2008a, b), he showed that a set of oscillations satisfying conditions (\ref{0}) are
resonantly excited when their wave energies have opposite signs, assuming that 
a Lindblad resonance occurs in the propagation region of the oscillations.
In a subsequent study by Kato et al. (2011) (hereafter, Paper I), it is shown that the presence of Lindblad resonance 
in wave propagation region is not essential for excitation. 
What is essential is opposite signs of wave energy of resonantly
interacting oscillations with a sufficient accuracy of $\omega_2\sim \omega_1\pm \omega_{\rm D}$.
Numerical calculations made by Ferreira and Ogilvie (2008), however, suggests that in addition to 
the above necessary conditions, a dissipative process in one of oscillations  
is necessary to excite the set of oscillations.
As the dissipative process, they consider the corotation damping.
Except for p-mode oscillations with no node in the vertical direction, 
non-axisymmetric oscillations are known to be damped by the presence of corotation resonance, even if the corotation 
point (radius) appears in the evanescent region of the oscillations.
To mimic this corotation damping, they take into account an artificial resistive force acting on oscillations in their numerical
calculations. 

In this paper we take into account a resistive force, as Ferreira and Ogilvie (2008) did, into 
the formulation, and examine how in our formulation the excitation of the set of possitive- 
and negative-energy oscillations is affected by the presence of the dissipative process.
The results of analyses show that the difference between Kato et al. (Kato 2004, 2008a, b:
Kato et al. 2011) and Ferreira and Ogilvie (2008) comes from seeing different aspects of the same resonant excitation process.

\section{Basic Hydrodynamical Equations and Resistive Damping}

We follow the same procedures as those in Paper I, except that in the present study a resistive force acting
on a fluid element is additionally considered.
This resistive force is assumed to work in perturbed states in the direction opposite to 
the Lagrangian velocity perturbation $\delta\mbox{\boldmath $u$}$.
In this case, the hydrodynamical equation describing perturbations over the equilibrium state is
\begin{equation}
   \delta\biggr[\frac{D\mbox{\boldmath $u$}}{Dt}+\nabla\psi+\frac{1}{\rho}\nabla p\biggr]
          =-\beta\delta\mbox{\boldmath $u$},
\label{2.1}
\end{equation}
where $\delta$ denotes the Lagrangian variation and $D/Dt$ is the Lagrangian time derivative along the perturbed flow $\mbox{\boldmath $u$}$,
and $\beta$ is the coefficient of the resistive force and taken to be a positive constant. 
Other notations in equation (\ref{2.1}) have their usual meanings.
We are interested in relativistic disks, but, for simplicity, the Newtonian formulation is adopted.
The effects of the relativity is mimiced by using the pseudo-Newtonian potential introduced by Paczy\'{n}ski
and Wiita (1980).

Now, we introduce the Lagrangian displacement vector, $\mbox{\boldmath $\xi$}$, and
express equation (\ref{2.1}) in terms of $\mbox{\boldmath $\xi$}$.
Since $\delta\mbox{\boldmath $u$}$ is related to the Lagrangian time derivative along the unperturbed flow 
$\mbox{\boldmath $u$}_0(\mbox{\boldmath $r$})$, 
i.e., $D_0/Dt$, by $\delta\mbox{\boldmath $u$}=D_0\mbox{\boldmath $\xi$}/Dt$, and $\delta$ and $D/Dt$ are
commutative (Lynden-Bell and Ostriker 1967), we have
\begin{equation}
    \delta\biggr(\frac{D\mbox{\boldmath{$u$}}}{Dt}\biggr)
    =\frac{D_0}{Dt}\delta\mbox{\boldmath $u$}
          =\frac{D_0^2}{Dt^2}\mbox{\boldmath $\xi$}.
\label{2.1'}
\end{equation}
Hence, equation (\ref{2.1}) is written as 
\begin{equation}
      \frac{D_0^2}{Dt^2}\mbox{\boldmath $\xi$}+\delta\biggr(\nabla\psi+\frac{1}{\rho}\nabla p\biggr)
          =-\beta\frac{D_0\mbox{\boldmath $\xi$}}{Dt}.
\label{2.2}
\end{equation}

A general expression for the right-hand side of equation (\ref{2.2}) in terms of the displacement vector $\mbox{\boldmath $\xi$}$
has been obtained by Lynden-Bell and Ostriker (1967) in the case where the perturbations are adiabatic and have small amplitudes.
The results show that $\rho_0(\mbox{\boldmath $r$})$ times the left-hand side of equation (\ref{2.2}) can be expressed as
\begin{equation}
    \rho_0{\partial^2\mbox{\boldmath $\xi$}\over\partial t^2}
         +2\rho_0(\mbox{\boldmath $u$}_0\cdot\nabla)
                 {\partial\mbox{\boldmath $\xi$}\over\partial t}
      +\mbox{\boldmath $L$}(\mbox{\boldmath $\xi$}),
\label{2.3}
\end{equation}
where $\partial/\partial t$ is the Eulerian time derivative, $\rho_0(\mbox{\boldmath $r$})$ is 
the density in the unperturbed state, and
$\mbox{\boldmath $L$}(\mbox{\boldmath $\xi$})$ is a linear
Hermitian operator (see in detail Lynden-Bell and Ostriker 1967). 
Detailed expression for $\mbox{\boldmath $L$}(\mbox{\boldmath $\xi$})$ is unnecessary here except that it is a Hermitian.

Now, we consider the case where perturbations are weakly non-linear.
In this case equation (\ref{2.2}) is written as
\begin{equation}
    \rho_0{\partial^2\mbox{\boldmath $\xi$}\over\partial t^2}
         +2\rho_0(\mbox{\boldmath $u$}_0\cdot\nabla)
                 {\partial\mbox{\boldmath $\xi$}\over\partial t}
      +\mbox{\boldmath $L$}(\mbox{\boldmath $\xi$}) 
      = -\beta\rho_0\biggr[\frac{\partial}{\partial t}+(\mbox{\boldmath $u$}_0\cdot\nabla)\biggr]
       \mbox{\boldmath $\xi$}
      +\rho_0\mbox{\boldmath $C$}(\mbox{\boldmath $\xi$},
                                                       \mbox{\boldmath $\xi$}),
\label{2.4}
\end{equation}
where $\mbox{\boldmath $C$}(\mbox{\boldmath $\xi$}, \mbox{\boldmath $\xi$})$ is the weakly non-linear term.
No detailed expression for $\mbox{\boldmath $C$}(\mbox{\boldmath $\xi$},\mbox{\boldmath $\xi$})$ is given here
[see Kato (2004, 2008a) for detailed expressions], but an important characteristics of 
$\mbox{\boldmath $C$}(\mbox{\boldmath $\xi$},\mbox{\boldmath $\xi$})$ is that we have commutative relations 
(Kato 2008a)
for an arbitrary set of $\mbox{\boldmath $\eta$}_1(\mbox{\boldmath $r$})$, 
$\mbox{\boldmath $\eta$}_2(\mbox{\boldmath $r$})$, and $\mbox{\boldmath $\eta$}_3(\mbox{\boldmath $r$})$, e.g.,
\begin{equation}
    \int\rho_0{\mbox{\boldmath $\eta$}}_1
                             \mbox{\boldmath $C$}
                          ({\mbox{\boldmath $\eta$}}_2, 
                           {\mbox{\boldmath $\eta$}}_3)dV
        =\int\rho_0{\mbox{\boldmath $\eta$}}_1
                             \mbox{\boldmath $C$}
                          ({\mbox{\boldmath $\eta$}}_3, 
                           {\mbox{\boldmath $\eta$}}_2)dV
        =\int\rho_0{\mbox{\boldmath $\eta$}}_3
                              \mbox{\boldmath $C$}
                           ({\mbox{\boldmath $\eta$}}_1, 
                            {\mbox{\boldmath $\eta$}}_2)dV,
\label{commutative}
\end{equation}
where the integration is made over the whole volume.
As shown later, the presence of these commutative relations leads to a simple expression 
of instability criterion.
We suppose that the presence of these commutative relations is a general property of conservative systems
beyond the assumption of weak nonlinearity. 

Equation (\ref{2.4}) is the basic equation to be treated in this paper, which is reduced to the one used in
Paper 1 in the limit of $\beta=0$.
Before examining the effects of the $\beta$-term on resonant excitation of oscillations, we briefly
check whether in the case of $\mbox{\boldmath $C$}=0$, the $\beta$-term of equation (\ref{2.4}) really leads
to damping of oscillations.
To do so, let us take the time-dependence of oscillations to be exp($i\omega t$).
Then, in the case of no damping (i.e., $\mbox{\boldmath $C$}=0$), equation (\ref{2.4}) becomes
\begin{equation}
    -\omega^2\rho_0\mbox{\boldmath $\xi$}
         +2i\omega\rho_0(\mbox{\boldmath $u$}_0\cdot\nabla)
                 \mbox{\boldmath $\xi$}
      +\mbox{\boldmath $L$}(\mbox{\boldmath $\xi$})
      +\beta\rho_0[i\omega+(\mbox{\boldmath $u$}_0\cdot\nabla)]\mbox{\boldmath $\xi$}=0.
\label{2.6}
\end{equation}
This equation is multiplied by the complex conjugate of $\mbox{\boldmath $\xi$}$ and integrated
over the whole volume of disks, assuming that $\rho_0$ vanishes on disk surface.
Taking the imaginary part of the resulting equation (using the facts that $\mbox{\boldmath $L$}$ 
and $i\rho_0(\mbox{\boldmath $u$}_0\cdot\nabla)$ are Hermitian operators), we have
\begin{equation}
     \Im \omega=\frac{1}{2}\beta.
\label{2.7}
\end{equation}
This shows that the $\beta$-term acts so as to dampen the oscillations as expected.

\section{Coupling of Two Oscillations Through Disk Deformation}

As in Paper I, we consider the case where two normal modes of disk oscillations, 
$\mbox{\boldmath $\xi$}_1(\mbox{\boldmath $r$}, t)$ and $\mbox{\boldmath $\xi$}_2(\mbox{\boldmath $r$}, t)$, 
resonantly couple through disk deformation, $\mbox{\boldmath $\xi$}_{\rm D}(\mbox{\boldmath $r$}, t)$.
Their sets of frequency and azimuthal wavenumber are denoted by ($\omega_1$, $m_1$), ($\omega_2$, $m_2$),
and ($\omega_{\rm D}$, $m_{\rm D}$) as mentioned before.
For example, ($\omega_1$, $m_1$) means that $\mbox{\boldmath $\xi$}_1(\mbox{\boldmath $r$}, t)$ is
proportional to exp[$i(\omega_1 t-m_1\varphi$)], where $\varphi$ is the azimuthal coordinate of the cylindrical cordinates
($r$, $\varphi$, $z$) whose origin is at the disk center and the $z$-axis is in the direction of disk rotation.
The resonant conditions are then given by equation (\ref{0}).
In general, through the nonlinear coupling term $\rho_0\mbox{\boldmath $C$}(\mbox{\boldmath $\xi$}, \mbox{\boldmath $\xi$})$
[see eq. (\ref{2.4})], other modes than $\mbox{\boldmath $\xi$}_1$ and $\mbox{\boldmath $\xi$}_2$ appear, and
their amplitudes as well as those of $\mbox{\boldmath $\xi$}_1$ and $\mbox{\boldmath $\xi$}_2$ become time-dependent.
The purpose here is to examine under what conditions the amplitudes of $\mbox{\boldmath $\xi$}_1$ and
$\mbox{\boldmath $\xi$}_2$ increase with time when they interact each other through a given disk deformation.
In order to examine this problem, we assume that normal modes of oscillations form a complete set, 
and expand the whole oscillations, $\mbox{\boldmath $\xi$}(\mbox{\boldmath $r$},t)$, resulting from the coupling
in the form:
\begin{equation}
   \mbox{\boldmath $\xi$}(\mbox{\boldmath $r$},t)=
    A_1(t)\mbox{\boldmath $\xi$}_1(\mbox{\boldmath $r$},t)
    +A_2(t)\mbox{\boldmath $\xi$}_2(\mbox{\boldmath $r$},t)
    +\sum_\alpha A_\alpha(t)\mbox{\boldmath $\xi$}_\alpha(\mbox{\boldmath $r$},t),
\label{3.2}
\end{equation}
where $\mbox{\boldmath $\xi$}_\alpha(\mbox{\boldmath $r$},t)$, which is proportional to exp$[i(\omega_\alpha t-m_\alpha \varphi)]$, 
is a normal mode other than $\mbox{\boldmath $\xi$}_1$ and $\mbox{\boldmath $\xi$}_2$, and $\sum_\alpha$ represents the summation 
over such modes.
In equation (\ref{3.2}) the disk deformation is not included, since it is assumed to have a large
amplitude so that its time variation can be neglected in the present study of excitation of small
amplitude oscillations.

Substitution of equation (\ref{3.2}) into equation (\ref{2.4}) leads to 
\begin{eqnarray}
  &&  \sum_{i=1,2}2\rho_0\frac{dA_i}{dt}\biggr[i\omega_i+(\mbox{\boldmath $u$}_0\cdot\nabla)\biggr]\mbox{\boldmath $\xi$}_i
       +\sum_{i=1,2}\beta_i\rho_0A_i[i\omega_i+(\mbox{\boldmath $u$}_0\cdot\nabla)]\mbox{\boldmath $\xi$}_i
                \nonumber   \\
  && +\sum_{\alpha}2\rho_0\frac{dA_\alpha}{dt}\biggr[i\omega_\alpha+(\mbox{\boldmath $u$}_0\cdot\nabla)\biggr]\mbox{\boldmath $\xi$}_\alpha 
     +\sum_\alpha\beta_\alpha\rho_0 A_\alpha[i\omega_\alpha+(\mbox{\boldmath $u$}_0\cdot\nabla)]\mbox{\boldmath $\xi$}_\alpha
                \nonumber   \\
  && = \sum_{i=1,2}\frac{1}{2}A_i\biggr[A_{\rm D}\biggr(\rho_0\mbox{\boldmath $C$}(\mbox{\boldmath $\xi$}_i,\mbox{\boldmath $\xi$}_{\rm D})
   +\rho_0\mbox{\boldmath $C$}(\mbox{\boldmath $\xi$}_{\rm D},\mbox{\boldmath $\xi$}_i)\biggr)
   +A_{\rm D}^*\biggr(\rho_0\mbox{\boldmath $C$}(\mbox{\boldmath $\xi$}_i,\mbox{\boldmath $\xi$}_{\rm D}^*)
   +\rho_0\mbox{\boldmath $C$}(\mbox{\boldmath $\xi$}_{\rm D}^*,\mbox{\boldmath $\xi$}_i)\biggr)\biggr] 
    \nonumber \\
  && +\sum_\alpha\frac{1}{2}A_\alpha\biggr[A_{\rm D}\biggr(\rho_0\mbox{\boldmath $C$}(\mbox{\boldmath $\xi$}_\alpha,\mbox{\boldmath $\xi$}_{\rm D})
     +\rho_0\mbox{\boldmath $C$}(\mbox{\boldmath $\xi$}_{\rm D},\mbox{\boldmath $\xi$}_\alpha)\biggr)
   +A_{\rm D}^*\biggr(\rho_0\mbox{\boldmath $C$}(\mbox{\boldmath $\xi$}_\alpha,\mbox{\boldmath $\xi$}_{\rm D}^*)
     +\rho_0\mbox{\boldmath $C$}(\mbox{\boldmath $\xi$}_{\rm D}^*,\mbox{\boldmath $\xi$}_\alpha)\biggr)
   \biggr].
\label{3.3}
\end{eqnarray}
In deriving this equation, such small terms as $d^2A_i/dt^2$ and $\beta_idA_i/dt$ have been neglected, since we are interested 
in slow secular evolutions of $A'$s due to the $\beta$-term and the coupling term.
On the right-hand side of equation (\ref{3.3}), the coupling terms that 
are not related to the disk deformation are neglected.
Furthermore, the coefficients of damping term, $\beta$, of different normal modes of oscillations has been distinguished by
attaching subscript.
This is because, as mentioned before, they are introduced to mimic the corotation damping, which 
depend on modes of oscillations.
The asterisk $*$ denotes the complex conjugate.

As discussed in Paper I, the normal modes of oscillations can be regarded approximately as a set of orthogonal 
functions.
In particular, in the vertically isothermal disks, orthogonality of normal modes approximately holds with the 
weight function $\rho_0(\mbox{\boldmath $r$})$ (Paper I, see also Okazaki et al. 1987).
That is, for two normal modes, $\mbox{\boldmath $\xi$}_i$ and $\mbox{\boldmath $\xi$}_j$, with moderately short radial wavelength
we have
\begin{equation}
     \langle\rho_0\mbox{\boldmath $\xi$}_i^*\mbox{\boldmath $\xi$}_j\rangle 
         = \langle\rho_0\mbox{\boldmath $\xi$}_i^*\mbox{\boldmath $\xi$}_j\rangle \delta_{ij},
\label{3.4}
\end{equation}
where $\langle\rho_0\mbox{\boldmath $\xi$}_i^*\mbox{\boldmath $\xi$}_j\rangle$ is the integration of     
$\rho_0\mbox{\boldmath $\xi$}_i^*\mbox{\boldmath $\xi$}_j$ over the whole volume of the disk, 
the subscripts, $i$ and $j$, being symbols distinguishing different modes.

Equation (\ref{3.3}) is now multipluied by $\mbox{\boldmath $\xi$}_1^*$ and integrated over the whole
volume.
Then, using the orthogonal relation and the resonant conditions (\ref{0}), we have an equation describing
time evolution of $A_1(t)$, which is (for detailed procedures, see Paper I) 
\begin{eqnarray}
   &&4i\frac{E_1}{\omega_1}\biggr(\frac{dA_1}{dt}+\frac{1}{2}\beta_1A_1\biggr)   \nonumber \\
   &&= A_2A_{\rm D}\hat{W}_{12}{\rm exp}(-i\Delta_+ t)\delta_{m_1,m_2+m_{\rm D}}
     + A_2A_{\rm D}^* \hat{W}_{12*}{\rm exp}(-i\Delta_- t)
        \delta_{m_1,m_2-m_{\rm D}},
\label{3.5}
\end{eqnarray}
where 
\begin{equation}
        \Delta_+=\omega_1-\omega_2-\omega_{\rm D} \quad {\rm and}\quad
        \Delta_-=\omega_1-\omega_2+\omega_{\rm D}.
\label{3.5'}
\end{equation}
The coupling terms ${\hat W}_{12}$ and ${\hat W}_{12*}$ in equation (\ref{3.5}) are time-independent quantities defined by\footnote{
It is noted that ${\hat W}_{12}$ and ${\hat W}_{12*}$ are time-independent quantities, since, for example, 
$\langle\rho_0\mbox{\boldmath $\xi$}_1^*\cdot
       \mbox{\boldmath $C$}(\mbox{\boldmath $\xi$}_2, \mbox{\boldmath $\xi$}_{\rm D})\rangle$
and $\langle\rho_0\mbox{\boldmath $\xi$}_1^*\cdot
       \mbox{\boldmath $C$}(\mbox{\boldmath $\xi$}_{\rm D}, \mbox{\boldmath $\xi$}_2)\rangle$
in equation (\ref{W12}) are proportional to ${\rm exp}(-i\Delta_+ t)$.  
}
\begin{equation}
       {\hat W}_{12}= \frac{1}{2}\biggr(\biggr\langle\rho_0\mbox{\boldmath $\xi$}_1^*\cdot
       \mbox{\boldmath $C$}(\mbox{\boldmath $\xi$}_2, \mbox{\boldmath $\xi$}_{\rm D})\biggr\rangle
       +\biggr\langle\rho_0\mbox{\boldmath $\xi$}_1^*\cdot\mbox{\boldmath $C$}(\mbox{\boldmath $\xi$}_{\rm D}, 
                 \mbox{\boldmath $\xi$}_2)\biggr\rangle\biggr){\rm exp}(i\Delta_+t),
\label{W12}
\end{equation}
and
\begin{equation}       
       {\hat W}_{12*}= \frac{1}{2}\biggr(\biggr\langle\rho_0\mbox{\boldmath $\xi$}_1^*
       \cdot\mbox{\boldmath $C$}(\mbox{\boldmath $\xi$}_2, \mbox{\boldmath $\xi$}_{\rm D}^*)\biggr\rangle
       +\biggr\langle\rho_0\mbox{\boldmath $\xi$}_1^*
       \cdot\mbox{\boldmath $C$}(\mbox{\boldmath $\xi$}_{\rm D}^*, \mbox{\boldmath $\xi$}_2)\biggr\rangle\biggr){\rm exp}(i\Delta_-t).
\label{W12'}
\end{equation}
In equation (\ref{3.5}), $E_1$ is the wave energy of mode 1 and defined by (e.g., Kato 2001)
\begin{equation}
   E_1=\frac{1}{2}\omega_1\biggr[\omega_1\langle\rho_0\mbox{\boldmath $\xi$}_1^*\mbox{\boldmath $\xi$}_1\rangle
         -i\langle\rho_0 \mbox{\boldmath $\xi$}_1^*(\mbox{\boldmath $u$}_0\cdot\nabla)\mbox{\boldmath $\xi$}_1\rangle\biggr],
\label{energy1}
\end{equation}
which is a real quantity, since $i\rho_0(\mbox{\boldmath $u$}_0\cdot\nabla)$ is a Hermitian operator.
The symbol $\delta_{a,b}$ in equation (\ref{3.5}) means that it is the delta-function, i.e., unity when $a=b$, 
while zero when $a\not= b$.

Similarly, by multiplying $\mbox{\boldmath $\xi$}_2^*$ to equation (\ref{3.3}) and integrating over the whole volume, 
we have an equation describing time evolution of $A_2(t)$, which is
\begin{eqnarray}
   &&4i\frac{E_2}{\omega_2}\biggr(\frac{dA_2}{dt}+\frac{1}{2}\beta_2A_2\biggr)   \nonumber \\
   &&= A_1A_{\rm D}\hat{W}_{21}{\rm exp}(i\Delta_- t)\delta_{m_2,m_1+m_{\rm D}} 
       + A_1A_{\rm D}^* \hat{W}_{21*}{\rm exp}(i\Delta_+ t)
        \delta_{m_2,m_1-m_{\rm D}},
\label{3.8}
\end{eqnarray}
where
$\hat{W}_{21}$ and $\hat{W}_{21*}$ are time-independent quantities defiend by 
\begin{equation}
       \hat{W}_{21}= \frac{1}{2}\biggr(\biggr\langle\rho_0\mbox{\boldmath $\xi$}_2^*\cdot
       \mbox{\boldmath $C$}(\mbox{\boldmath $\xi$}_1, \mbox{\boldmath $\xi$}_{\rm D})\biggr\rangle
       +\biggr\langle\rho_0\mbox{\boldmath $\xi$}_2^*\cdot\mbox{\boldmath $C$}(\mbox{\boldmath $\xi$}_{\rm D}, 
                 \mbox{\boldmath $\xi$}_1)\biggr\rangle\biggr){\rm exp}(-i\Delta_-t),
\label{W21}
\end{equation}
\begin{equation}
       \hat{W}_{21*}= \frac{1}{2}\biggr(\biggr\langle\rho_0\mbox{\boldmath $\xi$}_2^*\cdot
       \mbox{\boldmath $C$}(\mbox{\boldmath $\xi$}_1, \mbox{\boldmath $\xi$}_{\rm D}^*)\biggr\rangle
       +\biggr\langle\rho_0\mbox{\boldmath $\xi$}_2^*\cdot\mbox{\boldmath $C$}(\mbox{\boldmath $\xi$}_{\rm D}^*, 
                 \mbox{\boldmath $\xi$}_1)\biggr\rangle\biggr){\rm exp}(-i\Delta_+t),
\label{W21*}
\end{equation}
and $E_2$ is the wave energy of mode 2 defined by
\begin{equation}
   E_2=\frac{1}{2}\omega_2\biggr[\omega_2\langle\rho_0\mbox{\boldmath $\xi$}_2^*\mbox{\boldmath $\xi$}_2\rangle
         -i\langle\rho_0\mbox{\boldmath $\xi$}_2^*(\mbox{\boldmath $u$}_0\cdot\nabla)\mbox{\boldmath $\xi$}_2\rangle\biggr].
\label{energy2}
\end{equation}
In the limit of $\beta_1=\beta_2=0$, equations (\ref{3.5}) and (\ref{3.8}) are reduced, respectively, 
to equations (29) and (36) in Paper I.

Here, it is of importance to note that we have the following identical relations:
\begin{equation}
      {\hat W}_{21}=({\hat W}_{12*})^*,
\label{3.14}
\end{equation}
\begin{equation}
      {\hat W}_{21*}=({\hat W}_{12})^*.
\label{3.15}
\end{equation}
These relations come from the fact that for arbitrary functions, $\mbox{\boldmath $\eta$}_1$,
$\mbox{\boldmath $\eta$}_2$, and $\mbox{\boldmath $\eta$}_3$, their order in 
$\langle\rho_0\mbox{\boldmath $\eta$}_1\cdot\mbox{\boldmath $C$}(\mbox{\boldmath $\eta$}_2, \mbox{\boldmath $\eta$}_3)\rangle$
can be arbitrary changed [see eq. (\ref{commutative}) and Kato 2008a]. 

\section{Equation Describing Growth Rate}

Equation (\ref{3.5}) and (\ref{3.8}) show that in the limit of no coupling, $A_1$ and $A_2$ damp, respectively, by the
rate of $\beta_1/2$ and $\beta_2/2$, as expected.
The next problem is to examine how $A_1$ and $A_2$ evolve with time by the resonant coupling.
This is studied in the two cases of $m_2=m_1+m_{\rm D}$ and $m_2=m_1-m_{\rm D}$, separately.

\subsection{Case of $m_2=m_1+m_{\rm D}$}

New variables defined by
\begin{equation}
       {\tilde A}_1(t)=A_1(t){\rm exp}\biggr(i\frac{1}{2}\Delta_- t\biggr),
\label{4.1}
\end{equation}
\begin{equation}
     {\tilde A}_2(t)=A_2(t){\rm exp}\biggr(-i\frac{1}{2}\Delta_- t\biggr)
\label{4.2}
\end{equation}
are introduced.
Then, from equations (\ref{3.5}) and (\ref{3.8}), we have\footnote{
In Paper I, we have introduced ${\tilde A}_1(t)$ defined by ${\tilde A}_1(t)=A_1(t){\rm exp}(i\Delta_-t)$, different
from equation (\ref{4.1}).
The definition of ${\tilde A}_1(t)$ by equation (\ref{4.1}) is, however, better, since the resulting equations (\ref{4.3})
and (\ref{4.4}) are symmetric.
}, respectively,
\begin{equation}
     4i\frac{E_1}{\omega_1}\biggr(\frac{d}{dt}-i\frac{1}{2}\Delta_-+\frac{1}{2}\beta_1\biggr){\tilde A}_1={\tilde A}_2
          A_{\rm D}^*{\hat W}_{12*},
\label{4.3}
\end{equation}
and 
\begin{equation}
     4i\frac{E_2}{\omega_2}\biggr(\frac{d}{dt}+i\frac{1}{2}\Delta_-+\frac{1}{2}\beta_2\biggr){\tilde A}_2={\tilde A}_1
          A_{\rm D}{\hat W}_{21}.
\label{4.4}
\end{equation}
Equation ({\ref{4.3}) and (\ref{4.4}) are easily solved by taking ${\tilde A}_1$ and ${\tilde A}_2$ 
to be proportional to ${\rm exp}(i\sigma t)$.
The equation describing $\sigma$ is then found to be
\begin{equation}
   \sigma^2-i\frac{1}{2}(\beta_1+\beta_2)\sigma-\frac{1}{4}\Delta_-^2
        +i\frac{1}{4}\Delta_-(\beta_2-\beta_1)-\frac{1}{4}\beta_1\beta_2
       -\frac{\omega_1\omega_2}{16E_1E_2}\vert A_{\rm D}\vert^2\vert{\hat W}_{21}\vert^2=0,
\label{4.5}
\end{equation}
where equation (\ref{3.14}) has been used.

In the limit of $\beta_1=\beta_2=0$, equation (\ref{4.5}) is reduced to
\begin{equation}
    \sigma^2=\frac{1}{4}\Delta_-^2+\frac{\omega_1\omega_2}{16E_1E_2}\vert A_{\rm D}\vert^2\vert{\hat W}_{21}\vert^2,
\label{4.5'}
\end{equation}
and the oscillations are unstable ($\sigma^2<0$) when
\begin{equation}
      \Delta_-^2+\frac{\omega_1\omega_2}{4E_1E_2}\vert A_{\rm D}\vert^2\vert{\hat W}_{21}\vert^2<0.
\label{4.5''}
\end{equation}
This is the instability condition obtained in Paper I, and $(\omega_1\omega_2/E_1E_2)<0$ 
is a necessary condition for instability.
In another limit of $\beta_1=\beta_2\equiv \beta$, equation (\ref{4.5}) gives
\begin{equation}
   \biggr(\sigma-i\frac{1}{2}\beta\biggr)^2=\frac{1}{4}\Delta_-^2+
        \frac{\omega_1\omega_2}{16E_1E_2}\vert A_{\rm D}\vert^2\vert{\hat W}_{21}\vert^2.
\label{4.5'''}
\end{equation}
This equation shows that the term of $\beta$ always works so as to dampen the oscillations, since $\beta$ 
appears in the set of $\sigma-i\beta/2$.

An interesting and important effect of $\beta$, however, appears when $\beta_2\not= \beta_1$,
which will be considered in the next section.

\subsection{Case of $m_2=m_1-m_{\rm D}$}

In this case we introduce variables ${\tilde A}_1$ and ${\tilde A}_2$ defined by
\begin{equation}
       {\tilde A}_1(t)=A_1(t){\rm exp}\biggr(i\frac{1}{2}\Delta_+ t\biggr),
\label{4.6}
\end{equation}
\begin{equation}
     {\tilde A}_2(t)=A_2(t){\rm exp}\biggr(-i\frac{1}{2}\Delta_+ t\biggr).
\label{4.7}
\end{equation}
Then, from equations (\ref{3.5}) and (\ref{3.8}), we have, respectively,
\begin{equation}
     4i\frac{E_1}{\omega_1}\biggr(\frac{d}{dt}-i\frac{1}{2}\Delta_++\frac{1}{2}\beta_1\biggr){\tilde A}_1={\tilde A}_2
          A_{\rm D}{\hat W}_{12},
\label{4.8}
\end{equation}
and 
\begin{equation}
     4i\frac{E_2}{\omega_2}\biggr(\frac{d}{dt}+i\frac{1}{2}\Delta_++\frac{1}{2}\beta_2\biggr){\tilde A}_2={\tilde A}_1
          A_{\rm D}^*{\hat W}_{21*}.
\label{4.9}
\end{equation}
As in the case of $m_2=m_1+m_{\rm D}$, the time dependences of ${\tilde A}_1(t)$ and ${\tilde A}_2(t)$ 
are taken to be ${\rm exp}(i\sigma t)$.
Then, from equation (\ref{4.8}) and (\ref{4.9}) we have an equation describing $\sigma$ as
\begin{equation}
   \sigma^2-i\frac{1}{2}(\beta_1+\beta_2)\sigma-\frac{1}{4}\Delta_+^2
        +i\frac{1}{4}\Delta_+(\beta_2-\beta_1)-\frac{1}{4}\beta_1\beta_2
       -\frac{\omega_1\omega_2}{16E_1E_2}\vert A_{\rm D}\vert^2\vert{\hat W}_{12}\vert^2=0,
\label{4.10}
\end{equation}
where equation (\ref{3.15}) has been used.
This equation is the same as equation (\ref{4.5}) except that $\Delta_-$ and ${\hat W}_{21}$ in equation (\ref{4.5}) are now changed 
to $\Delta_+$ and ${\hat W}_{12}$, respectively.

\section{Numerical Calculations of Growth Rate}

Equations (\ref{4.5}) and (\ref{4.10}) have similar forms.
Considering this, we numerically solve the following equation:
\begin{equation}
     \sigma^2-i\frac{1}{2}\beta(1+\delta)\sigma-\frac{1}{4}\Delta^2+i\frac{1}{4}\Delta\beta(1-\delta)-\frac{1}{4}\beta^2\delta+W=0,
\label{5.1}
\end{equation}
where $W$ is an abbreviation of $-(\omega_1\omega_2/E_1E_2)\vert A_{\rm D}\vert^2\vert {\hat W}_{21}\vert^2$ 
or $-(\omega_1\omega_2/E_1E_2)\vert A_{\rm D}\vert^2\vert {\hat W}_{12}\vert^2$
and taken to be positive, since we are interested in the case of $(\omega_1\omega_2/E_1E_2)<0$.
Furthermore, $\beta$ and $\delta$ represent $\beta_2$ and $\beta_1/\beta_2$, respectively.
In the case of equation (\ref{4.5}), $\Delta$ in equation (\ref{5.1}) is $\Delta_-$, 
while it is $\Delta_+$ in the case of equation (\ref{4.10}).

As mentioned in the previous section, an interesting case is that one of $\beta_1$ and $\beta_2$ is negligibly small 
compared with the other one.
Such a case is really expected.
For example, let us consider the case where an axisymmetric g-mode oscillation ($m_1=0$) with one node in the
vertical direction ($n_1=1$) resonantly interacts with an one-armed g-mode oscillation ($m_2=1$) with two nodes 
in the vertical direction ($n_2=2$) through a warp ($m_{\rm D}=1$, $n_{\rm D}=1$).\footnote{
This is one of the cases numerically examined by Ferreira and Ogilvie (2008).
}
In this case, the resonant condition of $m_2=m_1+m_{\rm D}$ is satisfied.
Furthermore, the resonant condition of $\omega_1=\omega_2+\omega_{\rm D}$ can also be satisfied with $\omega_{\rm D}\simeq 0$.
This is because the propagation region of the former oscillation is $\omega^2<\kappa^2$ ($\kappa$ being the epicyclic
frequency), while that of the latter one is $(\omega-\Omega)^2<\kappa^2$, and these two propagation regions
overlap in relativistic disks, since $\kappa(r)$ has the maximum near $r=4r_{\rm g}$ ($r_{\rm g}$ being the Schwarzshild radius).
The former oscillation has no corotation damping ($\beta_1=0$), since it is axisymmetric, while the corotation
damping occurs in the latter oscillation ($\beta_2\not= 0$).
Hence, this case corresponds to $\delta=0$, and equation (\ref{5.1}) is reduced to 
\begin{equation}
       \sigma^2-i\frac{1}{2}\beta\sigma-\frac{1}{4}\Delta^2+i\frac{1}{4}\beta\Delta+W=0.
\label{5.2}
\end{equation}

The opposite case of $\beta_2=0$ and $\beta_1\not= 0$ may occur, if we consider a different set of oscillations.	
Such cases are formally involved in equation (\ref{5.2}) by changing the sign of $\Delta$.
Considering these situations, we solve equation (\ref{5.2}) for various set of parameters.
Dimensionless parameters we adopt are $\Delta/W^{1/2}$ and $\beta/W^{1/2}$.
If we obtain the dimensionless frequency, $\sigma/W^{1/2}$, whose imaginary part is negative, the set of oscillations considered are
amplified (overstable).

First, we consider the range of parameters to be examined to understand the characteristics of equation (\ref{5.2}).
In the case of $\beta=0$, equation (\ref{5.2}) gives $\sigma^2=\Delta^2/4-W$, and the sign of $\sigma^2$ changes at
$\Delta/W^{1/2}=\pm 2$.
Hence, the range of $-4<\Delta/W^{1/2}< 4$ will be enough to examine the basic behavior of equation 
(\ref{5.2}), since both cases of $\sigma^2<0$ (growing oscillation) and $\sigma^2>0$
(constant amplitude oscillations) are involved in this range.
The contours of constant value of $-\Im \sigma/W^{1/2}$ (i.e., a normalized growth rate of oscillations) in the 
parameter plane of the $\Delta/W^{1/2}$ -- $\beta/W^{1/2}$ are shown in figure 1 by solving equation (\ref{5.2}), i.e., $\delta=0$..

\begin{figure}
\begin{center}
    \FigureFile(80mm,80mm){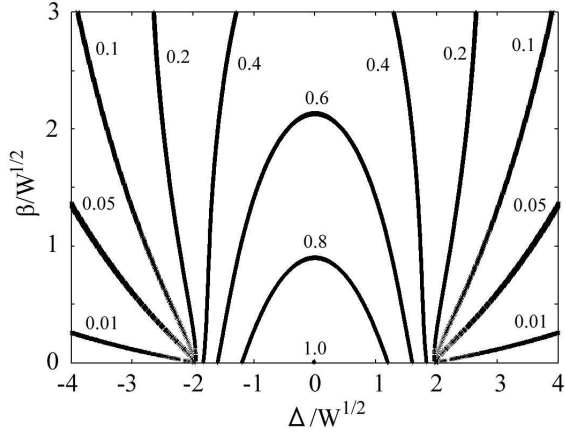}
\end{center}
\caption{Contours of constant (dimensionless) growth rate, $-\Im \sigma/W^{1/2}$, on the $\beta/W^{1/2}$ -- $\Delta/W^{1/2}$ plane.
The values labelled to curves denote dimensionless growth rate, $-\Im \sigma/W^{1/2}$.
The parameter $\delta$ is taken to be zero.} 
\end{figure}

Figure 1 shows that the growth rate of oscillations is the strongest at the point of $\Delta=0$ and $\beta=0$ on the
$\Delta/W^{1/2}$ -- $\beta/W^{1/2}$ diagram, and decreases as departing from the point on the diagram.
Next, let us see how the growth rate changes as $\beta$ increases from zero, keeping $\Delta$ fixed.
Roughly speaking, when $\vert\Delta\vert/W^{1/2}<2$, the growth rate decreases as $\beta$ increases.
In the case of $\vert\Delta\vert/W^{1/2}>2$, however, the situation is changed.
That is, in the latter case we have $\sigma^2=0$ and the oscillations are purely periodic
(there is no growth nor damping) in the limit of $\beta=0$, but the oscillations are amplified as $\beta$ increases from zero.
In other words, $\beta$ acts so as to excite oscillations.

In order to see more in detail the effects of $\beta$ on growth rate of oscillations, the $\beta$-dependence of
growth rate is shown in figure 2 for some values of $\delta$, fixing $\vert\Delta\vert/W^{1/2}$.
The figure shows that when $\delta=0$ and $\vert\Delta\vert/W^{1/2}>2$, the $\beta$ terms in equation (\ref{5.2}) 
act so as to excite oscillations.
If $\beta/W^{1/2}$ is sufficiently large, however, the growth rate tends to zero.

As mentioned before, the simultaneous presence of $\beta_1$ and $\beta_2$ acts so as to dampen the oscillations.
To demonstrate this numerically, the cases of $\delta=0.3$ and $\delta=1.0$ are added in figure 2, 
by solving equation (\ref{5.1}) [not equation (\ref{5.2})].

\begin{figure}
\begin{center}
    \FigureFile(80mm,80mm){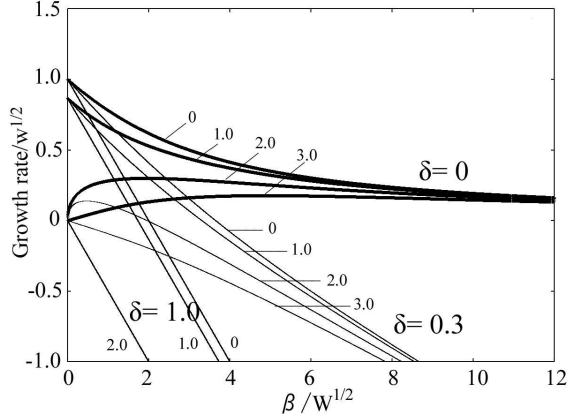}
\end{center}
\caption{Dimensionless growth rate, $-\Im \sigma/W^{1/2}$, as functions of parameter $\beta/W^{1/2}$ for
some values of $\Delta/W^{1/2}$ and $\delta$.
The thick curves are for $\delta=0$ (i.e., $\beta_1=0$), and the four curves are for $\Delta/W^{1/2} =
0$, 1.0, 2.0, and 3.0 from the top to the bottom.
The thin curves in the middle region of the figure are for $\delta=0.3$, and the values of 
$\Delta/W^{1/2}$ are 0, 1.0, 2.0, and 3.0 from the top to the bottom.
The three curves in the left-lower corner of the figure are for $\delta =1.0$, and the values of 
$\Delta/W^{1/2}$ are 0, 1.0, and 2.0 from the top to the bottom
(the curve for $\Delta/W^{1/2}= 3.0$ cannot be distinguished from that of $\Delta/W^{1/2}=2.0$ on the diagram).
} 
\end{figure}

\section{Summary of Growth Rate and Interpretation}

First, let us consider the case of no dissipation ($\beta=\delta=0$).
In this case the two oscillations are independent and have their own frequencies $\omega_1$ and $\omega_2$,
when their frequencies are far from the resonant conditions (\ref{0}).
Here, we examine only the case where the resonant conditions are $\omega_1\simeq \omega_2+\omega_{\rm D}$ and
$m_1=m_2+m_{\rm D}$ with $\omega_{\rm D}>0$, in order to avoid unnecessary complication considering other cases simultaneously.
If $\vert \Delta_+\vert$ decreases, for example, by a change of $\omega_{\rm D}$, the frequencies of two oscillation modes, 
$\omega_1$ and $\omega_2$, move toward $\omega_1-\Delta_+/2$ and $\omega_2+\Delta_+/2$, respectively, so that the 
exact resonant condition of frequencies is satisfied by these shifted frequencies, i.e., 
$(\omega_1-\Delta_+/2)=(\omega_2+\Delta_+/2)+\omega_{\rm D}$.\footnote{
For example, we have $\mbox{\boldmath $\xi$}_1\propto {\rm exp}(i\omega_1t)A_1\propto[i(\omega_1-\Delta_+/2)t]
{\tilde A}_1\propto {\rm exp}[i(\omega_1-\Delta_+/2+\sigma)t]$, where
$\sigma=(\Delta_+/2)(1-4W/\Delta_+^2)^{1/2}$ and tends to zero as $\Delta_+/W^{1/2}$ approaches $\pm 2$.
}
This is realized finally at $\vert\Delta_+\vert=2W^{1/2}$.
If $\vert\Delta_+\vert$ decreases further, the frequencies of oscillations no more change (frequency rocking), but the 
oscillations become overstable (amplified).
This is the results of Paper I.
The cause of this overstability is energy exchange between positive- and negative-energy oscillations by resonance.

In the case of $\vert\Delta_+\vert>2W^{1/2}$ with no dissipation, the oscillations are purely
time-dependent although frequencies are shifted from their proper ones, $\omega_1$ and $\omega_2$,
as mentioned above.
Now, we consider what happens if dissipative processes are added (i.e., $\beta$'s$\not= 0$) to the above
case of $\vert\Delta_+\vert>2W^{1/2}$.
The results show that the oscillations are amplified if a dissipative process works mainly  
on one of the oscillations (see figure 2).  
This result can be understood as follows.

Let us consider the behavior of the oscillation of $\mbox{\boldmath $\xi$}_1$ (mode 1).
As understood from the first paragraph of this section, two forces act on mode 1.
One is the restoring force (hereafter, force A) which acts so as to make mode 1 oscillate with its proper frequency $\omega_1$.
The other one is a force (hereafter, force B) which works so that a fluid element of the oscillation departs  
from the equilibrium state, i.e., works in the direction to make the oscillation unstable.
This latter force (force B) comes from the resonant coupling with the oscillation of mode 2
[see the term of the right-hand side of equation (\ref{4.8})].
In the case of $\vert\Delta_+\vert > 2W^{1/2}$, the former force (force A) is stronger than the latter one
(force B), and mode 1 is a constant amplitude oscillation when there is no dissipation.
If a dissipative force works on mode 2 (not mode 1), the force B working on mode 1 becomes
weak during an oscillation, since the resonant term depends on the amplitude of mode 2.
This means that the net restoring force  (i.e., force A minus force B) at a phase of oscillation
increases compared with that at the phase one cycle before due to the decrease of force B.
In other words, a fluid element associated with the oscillations returns to the equilibrium 
position with a faster velocity compared with that in one cycle before, and
the amplitude of oscillations increases with time.
If this amplification of mode 1 is stronger than the weakening of mode 2 by dissipative process,
the whole system of two oscillations of modes 1 and 2 are amplified.
This is the reason why dissipative process amplifies the set of oscillations of modes 1 and 2.

If a dissipative process works also on mode 1 (i.e., $\beta_1\not= 0$), the restoring force
acting on mode 1 (force A) decreases during one cycle of oscillation of mode 1, and thus the net restoring force 
acting on mode 1
decreases, leading to damping of the whole system of oscillations of modes 1 and 2.
It is noted that in the above arguments the roles of modes 1 and 2 can be changed. 

This excitation mechanism of oscillations is quite similar with double-diffusive instability 
known in other fields of astrophysics and oceanography.
For example, overstable convection due to double diffusion processes is known in magnetized or rotating media.
Let us consider, for instance, a stellar atmosphere where the temperature stratification is super-adiabatic but convection
is suppressed by restoring force of magnetic fields.
If thermal conduction works on fluid elements and is faster than diffusion process of magnetic field by magnetic diffusivity,
the net restoring force working on the fluid elements (magnetic restoring force minus buoyancy force)
increases during an oscillation, since the decrease of buoyancy force is faster than that of the magnetic restoring force.
Then, we can expect oscillations whose amplitude increases with time (overstable convection).
This mechanism of overstable convection is first recognized by Cowling (1958), although it is first found mathematically by
Chandrasekhar (see Chandrasekhar 1961).
Semi-convection also occurs in a chemically stratified medium by thermal conduction (Kato 1966), since 
chemical element diffusion is much slower than thermal diffusion.
All of them are excitation of overstable oscillations by double-diffusive instability.
As recent applications of double-diffusive instability, see, e.g., Charbonnel and Zahn (2007) and Zaussing and Spruit (2011)
for elements mixing in stars and Latter et al (2010) for protostellar disks. 

\section{Discussion}

In order to suggest one of possible origins of high-frequency quasi-periodic oscillations observed in NS  
and BH LMXBs, Kato (2004, 2008a, b) suggested by analytical calculations that in
deformed (warp or eccentric deformation) disks, a set of positive- and negative-energy oscillations can
be resonantly excited.
Ferreira and Ogilvie (2008) and Oktarian et al. (2010) made numerical calculations to examine whether the excitation 
really exists.
Subsequently, Kato et al. (2011) (Paper I) made an analytical examination by a method which is different
from Kato's original one (Kato 2004, 2008a, b).

The above analytical and numerical investigations now seem to conform that if positive- and negative-energy oscillations 
resonantly interact through disk deformation, they are excited.
However, there remains an important difference between the results by Kato's group and those by Ferreira and Ogilvie.
That is,  in the results by Ferreira and Ogilvie (2008), an additional ingredient is necessary for the excitation.
This is a dissipative process acting on one of the oscillations.
On the other hand, such dissipative process is unnecessary to excite the oscillations in analytical and semi-analytical  
calculations made by Kato and Oktariani et al. 

Our results in this paper seems to resolve this disagreement.
Our results in this paper show that when frequency deviation from the exact resonant condition is small,
i.e., $\vert\Delta\vert/W^{1/2}< 2$, the oscillations are excited without help of any dissipation process.
When the deviation is large, i.e., $\vert\Delta\vert/W^{1/2}> 2$, however, a dissipative process ($\beta\not= 0$) is
necessary to excite the oscillations.
In this context, we think that the case examined by Ferreira and Ogilvie (2008) corresponds to the case of
$\vert\Delta\vert/W^{1/2}>2$.
The following seems to also support this conjucture.
In the numerical results by Ferreira and Ogilvie (2008), the growth rate increases from zero with increase of 
the damping factor $\beta$, but saturates at a certain level (see figure 4 of their paper).
This is qualitatively consistent with our result shown in figure 2.\footnote{ In our results the growth rate 
decreases and tends to zero, when the damping factor increases further.
In figure 4 of Ferreira and Ogilvie (2008), such trend seems to be also present.
}

It is noted that the analyses made by Kato (2004, 2008a, b) and those made by Paper I and this paper
are rather different.
In the former studies, the presence of a Lindblad resonance in the propagation region of oscillations is assumed and
the condition of excitation of oscillations are derived by considering the behavior of oscillations around the resonance.
In the present study as well as in Paper I, such assumption is not involved in the analyses.
In this sense, the latter analyses are more general.
However, it should be rememberd that the magnitude of growth rate depends on the magnitude of the coupling terms such as 
$\vert {\hat W}_{12}\vert^2$ and $\vert{\hat W}_{21}\vert^2$, and they are large when a Lindblad resonance 
appears in the propagation region of the oscillations. 

It is also noted that for the present resonant excitation mechanism to work efficiently, the propagation regions of
the positive- and negative-energy oscillations must overlap in the radial direction.
Otherwise, the terms of resonant couplings are small and there is practically little excitation.
In nonself-gravitating disks, the overlapping of the propagation regions is easy to occur in relativistic disks.

In our excitation mechanism of disk oscillations, a disk deformation (e.g., warp or eccentric deformation) is necessary.
What kinds of origins of such disk deformation are conceivable?
There are some possibilities.
For example, the angular momentum axis of disks will not be always expected to align with the spin axis of
the central source.
Then, the disks are tilted and have precession.
In such disks, excitation of a set of disk oscillations with positive- and negative-energies are expected.
It is noted that the direct cause of this excitation is not the spin of the central source, but a disk deformation.
As some other possible causes of disk deformation, 
the disk instability due to irradiation from central sources (Pringle 1992) will be conceivable.
The Papaloizou-Pringle instability also makes the inner region of disks deformed at the phase where
the inner tori are formed (Machida and Matsumoto 2008).

Finally, let us briefly mention whether our excitation mechanism can describe the spin-frequency relation observed in QPOs of
millisecond pulsars.
When a burst occurs on the surface of spinning central neutron stars,
a non-axisymmetric pattern rotating with the spin frequency, $\omega_{\rm s}$, will be induced on the disks, 
i.e., $\omega_{\rm D}=\omega_{\rm s}$.
In such cases, excitation of oscillations which satisfy $\omega_1-\omega_2\sim\omega_{\rm s}$ is expected, 
and QPOs with such frequency relation are really observed (e.g., 4U 1702-43).
In some other millisecond pulsars, however, the observed frequency relation is 
$\omega_2-\omega_1\simeq \omega_{\rm s}/2$ (e.g., SAX J1808.4-3658).
At a glance, it seems to be difficult to describe the latter by the present resonant model, although
more considerations will be worthwhile.

\bigskip\noindent
The author thanks Atsuo T. Okazaki for discussions at the initial phase of this study.

\bigskip
\leftskip=20pt
\parindent=-20pt
\par
{\bf References}
\par
Chandrasekhar, S. 1961, {\it Hydrodynamics and Hydromagnetic Stability} (Clarendon Press, Oxford), 
      Chap. III, IV, and V\par
Charbonnel, C. \& Zahn, J.-P. 2007, A\&A, 467L, 15 \par
Cowling, T.G. 1958, {\it Magnetohydrodynamics} (Interscience Publ., New York), Chap.4\par
Ferreira, B.T. \& Ogilvie, G.I. 2008, MNRAS, 386, 2297 \par
Kato, S. 1966, PASJ, 18, 374\par
Kato, S. 2001, PASJ, 53, 1\par 
Kato, S. 2004, PASJ, 56, 905\par
Kato, S. 2008a, PASJ, 60, 111 \par
Kato, S. 2008b, PASJ, 60, 1387 \par
Kato, S., Fukue, J., \& Mineshige, S. 2008, Black-Hole Accretion Disks 
  -- Toward a New paradigm -- (Kyoto: Kyoto University Press)\par
Kato, S., Okazaki, A.-T., \& Oktariani, F. 2011, PASJ, 63, xxx (Paper I)\par
Latter, H., Bonart, J.F., \& Balbus S.A. 2010, MNRAS, 405, 1831\par
Lynden-Bell, D. \& Ostriker, J.P. 1967, MNRAS, 136, 293  \par
Machida, M. \& Matsumoto, R. 2008, PASJ 60, 613 \par
Okazaki, A.-T., Kato, S., \& Fukue, J. 1987, PASJ, 39, 457 \par
Oktariani, F., Okazaki, A.-T., \& Kato, S. 2010, PASJ, 62, 709 \par
Paczy\'{n}ski, B.,\& Wiita, P.J. 1980, A\&A 88, 23\par
Pringle, J.E. 1992, MNRAS 258, 811\par
Zaussinger, F. \& Spruit, H.C. 2011, astro-ph, arXiv: 1012.5851v2\par   
\leftskip=0pt
\parindent=20pt

\end{document}